# Stratified-turbulence observations in the deep Mediterranean

**by Hans van Haren**

Royal Netherlands Institute for Sea Research (NIOZ), P.O. Box 59, 1790  AB Den Burg,
the Netherlands.
e-mail: hans.van.haren@nioz.nl

**Abstract.** A nearly half-cubic hectometer of deep Mediterranean-Sea waters is yearlong sampled with about 3000 high-resolution temperature sensors to study different sources of turbulent waterflows, which are vital for life. Although temperature differences are never larger than 0.01°C, daily, weekly, and seasonal variations are observed. About half the time, relatively warm stratified waters are moved from 100's of meters higher levels to near the seafloor. These internal-wave and sub-mesoscale eddy-induced motions are half an order of magnitude more turbulent than those induced via general geothermal heating from below, and about one order of magnitude more turbulent than those from open-ocean processes. A rough estimate shows that eddy-induced stratified turbulence is likely more important for deep-sea life than rare, not observed, deep dense-water formation at the abyssal-plain mooring site. With a delay of about a week, the stratified turbulence tracks atmospheric disturbances, which are found 35% more energetic in winter than in summer. From comparison of turbulence-calculation methods, of band-pass filtering with vertical-displacement reordering, for data over one-four days, a generalization is proposed for the filter cut-offs under weakly stratified and near-homogeneous conditions in the deep Mediterranean.

# 1 Introduction

Irreversible, energy-consuming turbulence is indispensable for life on earth, also in the deep sea. Most ocean turbulence is generated near its boundaries, with an important input via the breaking of internal waves at steeply sloping underwater topography (Eriksen, 1982; Thorpe, 1987). Observational details are still scarce from the abyssal deep sea that is commonly considered to be 'quiescent and stagnant'. One of the key aspects in the breaking of internal waves is the 'warming phase', when, e.g., an internal tide moves downslope and characteristically (re-)stratifies waters near the seafloor with near-homogeneous waters due to convection turbulence above. Less known are slantwise-moving, warm turbulent waters possibly related with near-inertial waves and (sub-)mesoscale eddies in basins like the Mediterranean Sea where tides and stratified conditions are both weak. As will be demonstrated in this paper, these warm waters have the potential to generate larger turbulent mixing than in the open-ocean interior.

Under weakly stratified conditions, the potential of deep-sea turbulence generation via downward or slantwise moving waters should be compared with local general geothermal heating 'GH' through the seafloor (e.g., Pasquale et al., 1996), and with deep dense-water formation 'DWF' from the surface (e.g., Marshall and Schott, 1999). Sparse shipborne microstructure profiling has provided estimates of turbulence contributions from different sources across the Northwestern Mediterranean in which GH is found more important than internal wave breaking (Ferron et al., 2017). However, details of relevant processes are lacking and require high-resolution moored observations over prolonged periods of time.

In a stably-stratified environment like the sun-heated ocean, downward motions of warm water seem impossible in relation with irreversible convection turbulence involving motion in three spatial dimensions '3D'. Natural, body-forced buoyancy-driven convection (e.g., Dalziel et al., 2008; Ng et al., 2016) applies to denser waters moving down and less dense waters moving up in narrow plumes, i.e. in terms of temperature variations: warmer waters moving up and cooler waters moving down. In the ocean, such buoyancy-driven convection turbulence occurs regularly in the upper O(10) m near the surface during nighttime (e.g., Brainerd and Gregg, 1995) and possibly lower 100 m above the seafloor due to GH, depending on the local stratification. It can also occur as DWF after specific preconditioning of stratification reduction near the surface in localized areas like polar seas and the Mediterranean during brief irregular, rare periods (Marshall and Schott, 1999). An exemption can occur when the other major contributor to density variations dominates over temperature variations: if downward moving warm waters are sufficiently saltier than there environment, cooler and fresher waters may move up.

A reversible, also 3D, process occurs when internal-wave motions affect the stratified environment (e.g., LeBlond and Mysak, 1978). Such motions may displace relatively warm waters downward during a particular wave-phase, and cooler waters up. However, such displacements will not overturn and vertically mix the different water masses.

A combination of irreversible and reversible processes was observed in fresh-water alpine Lake Garda, where in the weakly stratified waters underneath internal waves convection turbulence was observed (van Haren and Dijkstra, 2021). As these observations showed similarity with the warming phase of a nonlinear wave breaking above a sloping seafloor (e.g., van Haren and Gostiaux, 2012), it was suggested that the convection underneath internal waves was either generated via shear moving convection tubes slantwise, or via wave-accelerations overcoming vertical density differences in internal-forcing overcoming reduced gravity, instead of body-forcing overcoming gravity as in natural convection. Similar to the effect of large-scale shear, planetary slantwise convection in the direction of the Earth's rotational vector may be brought about by the horizontal Coriolis parameter $f_h$ (Marshall and Schott, 1999). The convection can be induced via resonantly-forced standing inertial waves under homogeneous conditions (McEwan, 1973). At mid-latitudes, apparently-stable stratification having buoyancy frequencies of $N = f_h$, $2f_h$ or $4f_h$ occur in marginal stability due to planetary slantwise convection (van Haren, 2008).

In this paper, we further pursue the investigation of downward convective warm-water periods inducing stratified-turbulence that occur frequently in the deep Western Mediterranean. For this purpose, a nearly half-cubic-hectometer large 3D mooring-array is constructed holding about 3000 high-resolution temperature T-sensors and deployed at a deep seafloor. Turbulence calculations are made for a full observational year to investigate potential seasonal variations. Of the governing physics processes that indirectly may affect deep-sea life by inducing or transporting sufficient turbulence for nutrient and oxygen supply are: atmospheric-disturbance generated near-inertial waves, boundary-flow instability generating sub-, O(1) km, and meso-, O(10-100) km, scale eddies. Motions associated with these processes dominate dispersal of water masses in seas and oceans. While they are not considered to be part of irreversible 'small-scale' turbulent mixing, they are all greatly affected by the rotation of the Earth. How they transfer energy to small-scales of turbulence dissipation is not yet fully established. The large number of independent T-sensors is expected to improve statistics of turbulence values, in an environment where all dynamics is captured by temperature variations of less than 0.01°C. The small temperature variations put a large strain on the technical capabilities of quantifying the deep-sea turbulence.

## 2 Materials and Methods

In order to get some insight in the generation and development of deep-sea turbulence, 2925 independent, self-contained high-resolution NIOZ4 temperature 'T'-sensors were distributed in nearly half-a-million cubic meters of seawater. With two supplementary T-sensors registering tilt information above and below, 63 T-sensors were taped at 2-m intervals to 45 vertical lines 125-m tall that each were tensioned to 1.3 kN by a single buoy on top. The T-sensors were located between nominally h = 1.5-125.5±0.5 m above seafloor and recorded data at a rate of once per 2 s. Three buoys, of lines 1.4, 3.5 and 5.7 (henceforth throughout the text, the original naming 'group.line' is shortened without period; for layout see Appendix A1), held a single-point Nortek AquaDopp current meter 'CM' at h =126 m, which recorded data once per 600 s. The lines were attached at 9.5-m horizontal intervals to a steel-cable grid that was tensioned inside a 70-m diameter steel-tube ring functioning as a 140-kN anchor. The ensemble 'large-ring mooring' was deployed on the <1° flat and 2458-m deep seafloor of 42° 49.50′N, 006° 11.78′E just 10 km south of the steep continental slope, 5 km from its abyssal-plain foot, of the Northwestern Mediterranean Sea, in October 2020 (Fig. 1a). At the site's (mid-)latitude $f_h$ = 1.08f. Details of construction and deployment of the large-ring mooring can be found in van Haren et al. (2021). For calibration and reference purposes, a single shipborne Conductivity Temperature Depth 'CTD' profile was obtained to h = 0.5 m, about 1 km horizontally from the mooring site during the deployment cruise.

With the aid of Irish Marine Institute Remotely Operated Vehicle (ROV) "Holland I" all 45 vertical lines with T-sensors were successfully recovered in March 2024. Of the lines, 43 were mechanically in good order. Line 18 was hit by the drag parachute, which functioned as a stabiliser during the free-fall deployment, whereby 10 sensors were lost. Line 65 was about 0.5-m lower than nominal because of a loop near the cable grid. Fig. A1 shows the numbering of the lines, which were ordered in six groups for synchronisation purposes. As with previously deployed NIOZ4 T-sensors (for details see van Haren, 2018), the individual clocks were synchronised to a single standard clock every 4 hours, so that all T-sensors were sampled within 0.02 s. Line 36 did not register synchronization, possibly due to an electric cable failure. Three T-sensors leaked and <10 were shifted in position due to a tape malfunctioning. After calibration, some 20 extra T-sensors are not further considered due to electronics (noise) problems. In total, 2882 out of 2925 T-sensors functioned as expected for the first 20 months after deployment, with remaining bias due to electronic drift resulting in deviations from absolute accuracy. Depending on the period and type of analysis considered, between 50 and 150 T-sensors showed too large bias requiring additional attention during post-processing of the records from the weakly stratified deep sea.

Due to unknown causes all T-sensors switched off unintentionally when the file size on the memory card reached 30 MB. This may have to do with a formatting or programming error. It implied that a maximum of 20 months of data was obtained.

With respect to previous NIOZ4 T-sensor version, improvements of the electronics resulted in about twice lower noise levels of 0.00003°C and twice longer battery life. As described in van Haren (2018), calibration yielded a relative precision of <0.001°C. Bias due to instrumental electronic drift of < 0.001°C mo$^{-1}$ after aging was primarily corrected by referencing daily averaged vertical profiles, which must be stable from a perspective of turbulent overturning in a stratified environment, to a smooth polynomial without instabilities. In addition, because vertical temperature (density) gradients are so small in the deep Mediterranean, reference was made to periods of typically one hour duration that were homogeneous with temperature variations smaller than instrumental noise level (van Haren, 2022). Such periods were on days 350, 453, and 657 in the existing records. This secondary correction included low-pass noise filtering 'lpf' of data with time. Under near-homogeneous conditions, a tertiary correction involved lpf of data in the vertical. Temperature records were pressure-corrected by transferring to Conservative Temperature Θ (IOC et al., 2010) using CTD's mean local salinity value. Henceforth, Θ will be named 'temperature', for short.

Given the consistent and tight temperature-density relationship (Section 3), corrected temperature data allowed for calculations of turbulence values using the overturning displacement method of Thorpe (1977) by reordering density instabilities. Here, the method is applied under weakly stratified conditions in which buoyancy frequency $N \ll 10f$, f denotes the local inertial frequency. For the present deep-sea area, distinction is made between periods under environmental conditions when $N \sim > f$, somewhat exaggerating named stratified-water 'SW' conditions, and $N \sim < f$ including unstable values, named near-homogeneous 'NH' conditions. For NH, the tertiary correction is needed, and, when unstable overturns exceed the 124-m vertical range of sensors, an extra correction is mandatory because the Thorpe (1977) method of reordering is over-estimating displacements and resulting stratification (van Haren, 2025). Such periods are difficult to trace, because of the extremely small vertical temperature differences, and the selection can only be done manually as it is inadequately automated.

For comparison with mean Thorpe (1977) method, turbulence values are also computed using 'Ellison'-scales (Ellison, 1957 for atmospheric data; Itsweire, 1984 for laboratory data; Moum, 1996 for oceanographic microstructure profiler data). Such scales are determined from moored T-sensor time series by filtering out internal wave and sub-mesoscale motions. The method is quite sensitive for the precise high-pass filter 'hpf' cut-off frequency, as was noted for well-stratified Atlantic Ocean waters (Cimatoribus et al., 2014 for oceanographic moored T-sensor data). Here in Appendix A2, a modified version for application to moored T-sensor data under very weakly stratified conditions is proposed and filter cut-off frequencies are given for SW and NH conditions in the deep Mediterranean. For both methods, mean values are obtained from moored multiple-line T-sensor records after averaging over at least the largest turbulence scales, over the vertical '[…]', over time '<…>', and over 45 horizontally distributed lines '(…)'.

# 3 Results

The focus is on the first full year of observations to investigate potential seasonal variation in deep-sea turbulence values. The yearlong data-overview time series in Fig. 2 demonstrates 2-30 day variations, in waterflow speed (Fig. 2a), temperature (Fig. 2b), horizontal velocity difference (Fig. 2c), and vertical temperature difference (Fig. 2d). Such time-variability is typical for sub-meso- and mesoscale motions, which are likely associated with the dynamically unstable, meandering boundary current over the canyon-incised steep continental slope (Crepon et al., 1982) and which may develop into eddies.

Over the one year of observations, the waterflow speed U seldom exceeds 0.1 m s$^{-1}$, with little variations through the seasons. U also rather strongly varies with local inertial period, which partially reflects the variable thickness of the graphical curves in Fig. 2a.

Inertial motions do not dominate T-sensor data (Fig. 2b,d). Also in contrast with U, the T-data demonstrate a seasonal variation with relatively warmer (Fig. 2b) and more stratified (Fig. 2d) waters in winter, coarsely between days 365 and 495. The entire dynamical temperature variation over the year and up to h = 125 m from the seafloor is captured within maximum $|\Delta\Theta| < 0.01°C$, and commonly amounts only a few millidegrees. About half the time, $\Delta\Theta > 0.0002°C \equiv T_{thres}$, a threshold level that is about ssix times the standard deviation of T-sensor noise level and which provides an overall stratification resulting in N > 0.65f. These relatively warm SW either come from above or from the side. The other half of the time $\Delta\Theta < T_{thres}$, N < 0.65f and stratification may be unstable, or NH conditions. Under NH, less than 0.7% of total time negative temperature differences are found exceeding the (absolute value of) threshold level and corresponding with large-scale, >125-m developed GH. As a result, at the observational site convection turbulence associated with GH is suppressed by warm waters advected into the area most of the time.

Although the boundary current is strongest near the surface, it manifests itself at great depths including mesoscale variations at horizontal scales O(10-100) km. However, observed spatial waterflow variations over about 50-m horizontally indicate much smaller-scale, rapidly-fluctuating differences (Fig. 2c). These variations associate, in absolute value, with warm SW conditions in approximately half the cases. No cooling, inversely stratified waters, from above are observed in the yearlong record.

Compared to open-ocean waters where large 100-m scale $N > 10f$, the deep Mediterranean SW are characterized by weak stratification with $N = O(f)$. Despite the relatively weak stratification, SW will prove important for turbulent mixing in the area.

The single shipborne CTD observations show no dominant influence of salinity over temperature governing density variations, in the lower 500 m above the seafloor (Fig. 3). Over the well-resolved stratified portion between $-2165 < z < -2055$ m, the density-temperature relationship is found to be consistent (cf. van Haren, 2025),

$$\delta\sigma_2/\delta\Theta = -0.25\pm0.01 \text{ kg m}^{-3} \text{ °C}^{-1}, \tag{1}$$

where $\delta\sigma_2$ denotes the density anomaly referenced to a pressure level of $2\times10^7$ Pa. Hence, $\Theta$ can be used as tracer for density variations to quantify turbulent overturning using the Thorpe (1977) method.

In the weakly stratified waters over a vertical range of 100 m, density stratification varies, so that $N < 1f$ or NH is found near the seafloor, and $N \approx 2f$ or SW around $z = -2050$ m (Fig. 3d). Over 25-m vertical ranges, $N \geq 2f$ can be found (Fig. 3d), and over 1-10 m ranges $N > 4f$ may be inferred from Fig. 3c around $z = -2110$ and $-1980$ m. Such thin stratified layers are occasionally also found at greater depths if the better-resolved temperature profile is investigated (Fig. 3b). While the occasional vertical temperature differences of $<0.01$°C in Fig. 2d could result from horizontal differences or fronts, it seems more likely that stratification of around $z = -2050$ m in Fig. 3 is periodically lowered by action from above. Such action is expected down to about $h < 10$ m from the seafloor, a thin layer in which vertical temperature differences are generally very small but not always (black graph in Fig. 2d).

Thus, although the moored T-sensors were located between the seafloor and $z = -2332$ m, CTD-measured stratification may vary considerably with depth and time, and physical processes may lower warmer water some 400 m or advecting such waters slantwise, or possibly quasi-horizontally into the range of T-sensors. The precise direction of warm-water motion cannot be determined from single-station profiles, but may be resolved with a properly scaled 3D mooring-array.

### 3.1 1D-details of an arbitrary warm-water period

Considering an average 5-mK-amplitude warm-water period (cf. Fig. 2b), a 1.3-day depth-time detail from around day 485 is presented from single line 15 (Fig. 4). While waters seem depressed from h > 125 m, the warming occurs in variable periods of <1 hour (Fig. 4a). During the second half of the warming, 0.001°C additional heat is observed near the top. Relatively warm waters reach the seafloor twice within an inertial period of 0.73 day, around days 485.45 and 485.80. The warming ends with two cooler-water fronts and large overturning reaching the seafloor around day 486.2.

The warming is depressed to within h < 10 m from the seafloor, with a relatively large vertical temperature gradient between the cooler waters near the seafloor and the warmer waters higher up. This is reflected in the increased value of 2-m-small-scale buoyancy frequency $N_s$ in h < 30 m. Turbulent overturns hardly occur between days 485.2 and 486.15 for h < 5 m, but are non-negligible in the stratified waters above for 5 < h < 30 m, and are typically 50-m large further up for h > 30 m (Fig. 4b).

Quantifying turbulence dissipation rate requires averaging, over all overturning scales possible, and 124-m vertically averaged values demonstrate variations with time over two orders of magnitude, when reordered data are used (Thorpe (1977) method, black graph in Fig. 4c), or using 48 < ω < 3000 cpd (cycles per day) band-pass filtered data (Ellison (1957) method cf. Appendix A2, cyan graph in Fig. 4c).

Time-depth mean values for line 15 from SW's day 485 are: turbulence dissipation rate $<[\varepsilon_T]>$ = $6\pm3\times10^{-10}$ m$^2$s$^{-3}$ and turbulent diffusivity $<[K_z]>$ = $1.5\pm0.7\times10^{-3}$ m$^2$s$^{-1}$ under buoyancy frequency $<[N]>$ = $2.9\pm0.3\times10^{-4}$ s$^{-1}$ ≈ 3f, using Thorpe (1977) method. Modified Ellison (1957)-method $<[\varepsilon_E]>$ = $7\pm3\times10^{-10}$ m$^2$s$^{-3}$ (Appendix A2). These turbulence values are more than one order of magnitude larger than open-ocean values observed in stratified waters well away from boundaries (e.g., Gregg, 1989; Polzin et al., 1997; Yasuda et al., 2021).

### 3.2 1D-details of an arbitrary near-homogeneous period

For comparison, such turbulence values are under NH conditions between days 316.5 and 320.5 (Fig. 5): $2\pm1\times10^{-10}$ m$^2$s$^{-3}$ and $1.1\pm0.5\times10^{-2}$ m$^2$s$^{-1}$ under buoyancy frequency $<[N]>$ = $0.5\pm0.2\times10^{-4}$ s$^{-1}$ = 0.5f. These values follow partially correcting the original method by Thorpe (1977) for overturns exceeding the height of instrumentation with information from manually selected NH environments that are bounded by stratification above (van Haren, 2025). For periods with NH bounded by stratification above, such as between days 318.6 and 318.96, their mean turbulence dissipation rate is to within 10% the same as found for periods with convection turbulence due to GH: $\varepsilon_{GH} = 1.2 \times 10^{-10}$ m$^2$s$^{-3}$. The $\varepsilon_{GH}$ matches average geophysical heat-flux observations in the area (Pasquale et al., 1996), under the condition that the mixing coefficient of $\Gamma_C = 0.5$ (van Haren, 2025), which is typical for buoyancy-driven convection turbulence (Dalziel et al., 2008). For this period, $<[\varepsilon_E]> = 1.3 \pm 1 \times 10^{-10}$ m$^2$s$^{-3}$ (Appendix A2).

### 3.3 Some 45-line statistics of short periods under SW and NH conditions

For consistency and statistics, six half-day periods are considered for computation of turbulence dissipation rate values, three under SW and three under NH conditions. The computations are performed for all 45 vertical lines and averages are computed over the 124-m height and half-day periods. It provides a one-and-a-half order of magnitude distribution of mean turbulence dissipation rate values (Fig. 6).

While some values are highly consistent between lines, e.g. the most energetic period on day 441, others show a half-order of magnitude distribution of values like on day 459. Initially, this calculation was set-up to help identify biased T-sensors and the appropriate polynomial correction. After applied tertiary correction, remaining wide distributions are attributed to more general turbulence variability.

The statistics certainly improve turbulence dissipation rate values calculated using other instrumentation and methodology, which is generally to within a factor of two at best. The six examples of 45 lines provide about four times better statistics for the half-day periods (Table 1). The three NH values average to $<[\varepsilon_T]>_{NH} = 1.1 \pm 0.2 \times 10^{-10}$ m$^2$s$^{-3}$, which is well within error equivalent to $\varepsilon_{GH}$ for $\Gamma_C = 0.5$, while significantly different from the value for $\Gamma_C = 0.2$. It thus confirms previous results (van Haren, 2025) and laboratory findings for convection-turbulence (e.g., Dalziel et al., 2008). The three SW values average to $<[\varepsilon_T]>_{SW} = 8 \pm 5 \times 10^{-10}$ m$^2$s$^{-3}$, noting that the standard deviation of individual mean values are one order of magnitude smaller.

Another consequence of the use of multiple mooring lines besides improved statistics, is some insight in possible distribution of mean turbulence values. While one would expect erratic distribution over the short horizontal distances <70 m, particularly NH distributions yield two-dimensional consistent images such as on days 459 and 495 (Fig. 7). It provides confidence in consistency of methods used, but results in a puzzling gradient in turbulence that apparently is independent of waterflow (measured at h = 126 m). For these GH- and near-inertial eddies-dominated periods the 9.5-m interval between lines seems reasonably well chosen, where 100 m may have been too large.

**3.4 Yearlong daily averaged turbulence for 45 lines**

A yearlong time series of daily-averaged turbulence values is computed for all 45 lines (Fig. 8). This computation is automated, using a fixed $3^{rd}$-order polynomial for primary correction. Since the tertiary correction for >125-m extending overturns is not applied manually, a criterion for excluding such episodes is applied. The criterion is simply based on the daily-averaged temperature difference between uppermost and lowest T-sensor, per line. When $\Delta\Theta < T_{thres}$, given previously, the daily and vertical mean dissipation rate is fixed to,

$$<[\varepsilon_T]> = \varepsilon_{GH} = 1.2\times10^{-10} \text{ m}^2 \text{ s}^{-3}, \tag{3}$$

the mean value for geothermal heating (van Haren, 2025). This is found to occur in 59±1.5% of the time, somewhat varying per line, and characterizes NH, besides GH. About 40±1.5% of the time is characterized by SW.

The overall, yearlong, 125-m vertical, and 45-line mean turbulence dissipation rate amounts,

$$(<[\varepsilon_T]>) = 2.4\pm0.2\times10^{-10} \text{ m}^2 \text{ s}^{-3}, \tag{4}$$

so that the mean SW turbulence dissipation rate amounts,

$$(<[\varepsilon_T]>)_{SW} = 4.3\pm0.4\times10^{-10} \text{ m}^2 \text{ s}^{-3}, \tag{5}$$

which is thus closely represented by the short periods of days 308 and 485 in Fig. 6, Table 1.

Part of the SW-turbulence is attributable to GH in a layer of typically h = 30 m under stratified waters. This may be inferred from the vertical temperature difference in that layer that passes $T_{thres}$ during only 8% of daily periods, cf. the magenta graph in Fig. 2d. Nevertheless, the advection of warmer waters suppresses GH-turbulence, possibly affecting the small-scale distribution in Fig. 7, and the associated 3.5-times larger turbulence dissipation rates (5) are induced by convection and more generally by shear following internal-wave breaking.

Considering the yearlong 'seasonal' variation that was suggested from the temperature (difference) time series in Fig. 2, and which is represented by the logarithm of daily and vertically averaged <[N]> in Fig. 8c, the corresponding plot of <[$\varepsilon_T$]> (Fig. 8b) is more difficult to interpret, also in conjunction with meteorological data (Fig. 8a). Different-line data mostly collapse on each other during winter between days 365 and 495, for both <[N]> and <[$\varepsilon_T$]>. During this period, vertical-line daily mean turbulence dissipation rates most, 27 out of 41, exceed twice the mean value (4), shown by the green asterisks in Fig. 8a. The 11% of time of green-asterisks occurrence average to a mean turbulence dissipation rate of $7 \times 10^{-10}$ m$^2$ s$^{-3}$.

The average turbulence dissipation rate for the 130-day winter period is 25% higher than the yearlong mean (4). During this period, the wind work ~W$^2$ is increased by 20% compared to its yearlong average value. A rough visual correspondence is found between |W| (Fig. 8a) and lg<[$\varepsilon_T$]> (Fig. 8b), the former leading the latter by about one week. The clearest value-collapse of turbulence and stratification data from different lines is found between days 450 and 500. The early-spring period is unlikely governed by deep DWF due to limited meteorological forcing (Fig. 8a), but the preceding winter cooling may induce enhanced sub-mesoscale activity. Although increased sub-mesoscale motions can obscure near-inertial internal waves (van Haren and Millot, 2003), the transfer of energy to internal-wave scales leading to breaking and turbulence is not hampered. Possibly, as near-inertial shear is dominant in well-stratified waters, a shift from shear to convection turbulence may be associated with the increase of sub-mesoscale activity. Such potential energy transfer will be elaborated elsewhere.

In contrast, 14% lower mean turbulence dissipation rate than (4) is found during the summer between days 540-670. In this period, W$^2$ is decreased by 13% compared to its yearlong average value.

## 4 Discussion and conclusions

The observations show short O(10) day periods of typically 0.005°C warmer waters than their environment appearing from above, but also, as inferred from the 3D mooring-array, from the sides. The periods occur at a coarse near-inertial periodicity, which is much less deterministic than a tide, and at twice the inertial periodicity. Like internal waves in the Atlantic and Lake Garda (van Haren and Dijkstra, 2021), they push stratification to within a few meters from the seafloor. The pushdown is vigorously turbulent, more than one order of magnitude larger than in the open ocean away from boundaries. This relatively large turbulence should not surprise as both the bulk Reynolds number O($10^6$) and buoyancy Reynolds number $\varepsilon/(\nu N^2) = $ O($10^4$) are large, even in the weakly stratified deep sea. The $\nu \approx 10^{-6}$ m$^2$ s$^{-1}$ denotes the kinematic viscosity. As GH is found to be relatively weaker, the convection turbulence seems to be driven by the, slanted, internal waves from above.

While vertical motions by wintertime DWF have been observed via moored observations (e.g., Schott et al., 1996) and floats (Steffen and D'Asaro, 2002), and surface buoyancy fluxes have been estimated to be O($10^{-7}$) m$^2$ s$^{-3}$ during convection events (Marshall and Schott, 1999), quantification via observations of turbulence values associated with DWF reaching the abyssal seafloor have yet to be made (Thorpe, 2005). Unfortunately, a dense-water event never reached the large-ring mooring while it was underwater. Estimates of deep-convection duration are limited, albeit that some consensus exists about decadal variability or occurrence of seafloor-reaching DWF over a relatively short period of (less than) a week (Lilly et al., 1999), maximum a month, per 8-10 years (Dickson et al., 1996; Mertens and Schott, 1998). It is tempting to compare coarse DWF turbulence estimates with GH and SW turbulence calculated from observations at the present mooring site.

Because of the lack of measurements to quantify DWF turbulence, some insight is gained from nocturnal convection-turbulence near the ocean surface. Microstructure measurements by, e.g., Brainerd and Gregg (1995) demonstrated turbulence dissipation rate values > $10^{-7}$ m$^2$ s$^{-3}$ close to the surface and which decreased in the O(10) m near-homogeneous layer to typically,

$$\varepsilon_{DWF} \approx 10^{-8} \text{ m}^2 \text{ s}^{-3}, \tag{6}$$

at a depth just above well-stratified waters below. The one order of magnitude reduced value reflects the erosion of the stratification. Here, we take value (6) as a proxy for dissipation rate by an event of DWF-convection turbulence in waters just above the deep seafloor.

In comparison with GH's value (3), DWF's (6) is two orders of magnitude larger. Where GH is quasi-permanent, DWF rarely occurs, for example not at all during the presented 20 months of observations. Two orders of magnitude difference implies occurrence of (6) during one month per 8 years to match (3). This is the estimated maximum at a given site.

In comparison with mainly sub-mesoscale and internal-wave induced year-average value (4), DWF's (6) would have to occur during 2.5 months per 8 years, or during 9 days every year. This is not observed in the open Liguro-Provençal basin. It implies that, either DWF turbulence is stronger than (6) also for $z < -2000$ m, which seems unlikely, or GH turbulence and especially SW turbulence are several times, SW turbulence up to one order of magnitude, larger than DWF turbulence, when averaged over a decade in time. With their sources of sub-mesoscale eddies and near-inertial waves, the warm SW conditions thus seem more important than DWF for supply of fresh materials in the deep-sea area. Recall that the observations are made in an area where tides, normally about half the ocean's mechanical energy source, are weak.

The observed yearlong mean turbulence dissipation rate value of SW being 2.5 times that of GH in the present area is the reverse of findings by Ferron et al. (2017), who find three times larger GH than SW from sparse microstructure profiling across the entire Northwest Mediterranean. The discrepancy may have to do with the location of the large-ring mooring, about 5 km from the foot of the continental slope and most likely under the well-stratified boundary current most of the time.

Estimating turbulence dissipation rates from the microstructure profiler plots in Ferron et al. (2017) gives average values for h = 100-600 m (the instruments were stopped some 90 m above the seafloor) of about $2.5 \times 10^{-10}$ $m^2$ $s^{-3}$ and $7 \times 10^{-10}$ $m^2$ $s^{-3}$, for the Western Mediterranean and specifically Ligurian Sea, respectively. These values are in the same range as mean and SW values (4) and (5), respectively. Both averaged microstructure profiles showed reduction in values in the lower h = 100-200 m to about $1 \times 10^{-10}$ $m^2$ $s^{-3}$ and $3 \times 10^{-10}$ $m^2$ $s^{-3}$, respectively. For properly processed microstructure profiler data the instrumental error is to within a factor of two for mean turbulence dissipation rate values, not considering environmental variations. The shown values, from the same height above seafloor as the upper range of the moored T-sensors, compare with the GH-(3) and mean-(4) values determined at the large-ring mooring.

A contribution of salt to density variations may possibly affect the turbulence values calculated from the moored T-sensors under SW conditions, but density-temperature relationship across stratified layers is found consistent between different years (van Haren, 2025). Also, convection turbulence under SW conditions has been observed in deep alpine-lake Garda where salt contributes little to density variations (van Haren and Dijkstra, 2021).

The relative importance of stratified turbulence occurring in varying strength over about half the time has consequences for deep-sea transport, redistribution of matter and life. The regular replenishment is partially related with atmospheric disturbances, in an indirect way. Winds do not directly affect motions near the 2500-m deep seafloor. However, wind-induced near-inertial internal waves and boundary-current variations affecting sub-mesoscale eddies seem to have correspondence with turbulence intensity variations close to the seafloor, roughly a week after variations occur near the surface. More SW activity and about 20% larger turbulence dissipation rates were found in winter when atmospheric activity was correspondingly larger. Weakest stratification was found more in summer. As eddies and near-inertial waves cause convection in the direction of the rotational axis to slant to the vertical under weak z-direction stratification N = O(f), cf. McEwan (1973); Straneo et al. (2002); Sheremet (2004); Gerkema et al. (2008), the turbulence may come from above and in part horizontally.

In between stratified turbulence periods, waters tend to become near-homogeneous whilst being dominantly mixed by convection turbulence through GH. As in Rayleigh-Taylor convection, plumes of GH-response in waters overlying the seafloor strongly vary with time, and thus spatially (e.g., Dalziel et al. 2008; Ng et al., 2016).

Geologically, even over a nearly flat sedimented seafloor underlying crustal cracks may develop variable GH over distances as small as <1 km, depending on location of faults (e.g., Kunath et al., 2021). This could explain observed variation in mean turbulence dissipation rates over the 70-m size of the large-ring mooring, during GH. The 0.6-m high steel tubes of the large ring will not affect GH up to h

= 125 m. An inconclusive variation of mean turbulence values over periods of SW conditions demonstrates larger scale variability.

Overall, the 45 vertical lines and nearly 3000 high-resolution T-sensors provided improved statistics for daily mean turbulence dissipation rate values to within a reduced relative error of about 25%. Other strengths of the mooring-array like improved spectral resolution and 3D evolution of turbulence will be reported elsewhere, while short movies of 3D turbulence passages have been described in van Haren et al. (2026).

*Data availability.* Only raw data are stored from the T-sensor mooring-array. Analyses proceed via extensive post-processing, including manual checks, which are adapted to the specific analysis task. Because of the complex processing the raw data are not made publicly accessible. Current meter and CTD data are available from van Haren (2025): "Large-ring mooring current meter and CTD data", Mendeley Data, V1, https://doi.org/10.17632/f8kfwcvtdn.1. Atmospheric data are retrieved from https://content.meteoblue.com/en/business-solutions/weather-apis/dataset-api.

*Competing interests.* The author has no competing interests.

*Acknowledgments.* This research was supported in part by NWO, the Netherlands organization for the advancement of science. Captains and crews of R/V Pelagia are thanked for the very pleasant cooperation. NIOZ colleagues notably from the NMF department are especially thanked for their indispensable contributions during the long preparatory and construction phases to make this unique sea-operation successful. I am indebted to colleagues in the KM3NeT Collaboration, who demonstrated the feasibility of deployment of large three-dimensional deep-sea research infrastructures. E. Berbee, P. Kooijman. E. de Wolf, and E. Koffeman showed steep learning curves.

**Appendix A1 Layout of large-ring mooring**

The large-ring mooring has a diameter of nearly 70 m (Fig. A1). The eighteen 12-m long and 0.6-m diameter steel tubes hold a steel-cable grid for rigidness. The cables are 9.5 m apart. At cable-intersects, 2.5-m diameter 'small' rings are mounted that each held a 125-m long mooring line with 65 T-sensors below a single 1.45-kN buoy. Of eight small rings, imaginary intersects were at the steel tubes, so that special off-set mounting was needed with three assist cables (van Haren et al., 2021). Upon landing at the seafloor following parachute-controlled 'free' fall, the orientation of the ring was directed to the NNW, pointing at 337 °N. After underwater chemical release of the buoys, the cable-grid was lifted in a dome with its center h = 2.0 m above seafloor (van Haren, 2026 submitted).

The vertical mooring lines were named in six synchronisation groups, of maximum eight lines each. The single synchroniser was located at the small ring of central line 51. Every half hour, the synchroniser sent a clock pulse to a group. The synchronisation sequence of six pulses was repeated every four hours.

**Appendix A2 Proposed generalization of filter cut-off to compute Ellison scales**

Data from moored strings of high-resolution temperature sensors are potentially useful to compare two different manners of calculating turbulence values. The more common method proposed by Thorpe (1977) involves the reordering or sorting of unstable density overturns and the bookkeeping of their vertical displacements, for vertical profiles at each time step. Turbulence values are computed following averaging in the vertical, in time, or both and should include the largest of overturn scales. The method requires a consistent temperature-density relationship. In near-homogeneous waters, difficulties may arise in establishing such a relationship, but also in determining the size of largest overturns when they outgrow the height of the string. A method is proposed to correct over-estimation under convection turbulence by GH using verification via results from geophysical sampling (van Haren, 2025). In well stratified waters, the Thorpe (1977) method has been successfully compared (e.g., Itsweire, 1984; Moum, 1996; Cimatoribus et al., 2014), with results from the method introduced for atmospheric data by Ellison (1957). In this Appendix, such a comparison is done for weakly stratified waters. A modification is proposed of the Ellison (1957) method for application to data from moored instrumented strings, and which is practically based on instrument performance and environmental physics conditions.

Ellison (1957) separated time series of potential temperature $\theta(t, z)$, which is dynamically equivalent to Conservative Temperature $\Theta$ (IOC et al., 2010) in the ocean, at a fixed vertical position z in two,

$$\theta = <\theta> + \theta',  \tag{A1}$$

where $<.>$ denotes the lpf series and the prime its hpf equivalent. If multiple sensors are deployed in the vertical to establish a mean vertical gradient, a scale height can be defined as,

$$L_E = <\theta'^2>^{1/2}/(d<\theta>/dz).  \tag{A2}$$

Itsweire (1984), using laboratory CTD-profile data, and Moum (1996), using ocean microstructure profiler data, apply sorting as filter in z-direction. While their quasi-hpf data unlikely contain (linear) internal waves, provided the profiles were instantaneously made and strictly vertical, they may contain instrumental noise. With limited time-evolution available, it is assumed that the hpf data have worked against the local stratification. The same assumption is made for Thorpe (1977) displacements. Sorting works on all scales of overturns, which can be highly varying.

Like Thorpe-displacement 'd' scales, the Ellison scales of (A2) may be compared with the Ozmidov (1965) scale $L_O = cL_E$ of largest possible turbulent overturns in stratified waters, so that the turbulence dissipation rate reads,

$$\varepsilon_E = c^2 L_E^2 N^3, \tag{A3}$$

in which the constant c needs to be established. If we take an average value of c = 0.8 (Dillon, 1982), like commonly used for vertical root-mean-square 'rms' Thorpe scale $L_T = [d^2]$ so that $\varepsilon_T = c^2 L_T^2 N^3$, one can compare average dissipation rate values between the two methods.

While the Thorpe (1977) method is most sensitive to proper resolution of the largest vertical overturn scale, and the stratification (or buoyancy frequency) it works against, determination of Ellison scales from moored-sensor time series is most sensitive for the appropriate separation between internal waves and turbulent motions (Cimatoribus et al., 2014). For data from instrumented strings under well-stratified Northeast-Atlantic conditions, wavelet decomposition worked using an averaging scale of about 2/N for the lpf in (A1). Under such conditions, instrumental flaws like short-term bias of T-sensor data were minimal. However, such a determination of Ellison scales is not a straightforward task under weakly stratified and near-homogeneous conditions like occurring in the deep Western Mediterranean.

First, because such conditions imply very small variations in temperature (density), time series require lpf to remove instrumental noise.

Second, time series require hpf to remove internal waves and (sub-)mesoscale motions. While the common internal-wave band is considered between ranges f and N, for well stratified waters N >> f, more complex inertio-gravity wave 'IGW' frequency 'ω' bounds [$\omega_{min}$<f, $\omega_{max}$>N], for large-scale mean N, have to be considered in waters where N = O(f), e.g. (LeBlond and Mysak, 1978; Gerkema et al., 2008). Furthermore, while average large-scale stratification hampers turbulent overturns, internal-wave straining separates small thin-layer stratification, with small-vertical-scale buoyancy frequency $N_s$, from near-homogeneous layers, with minimum buoyancy frequency $N_{min}$, which may carry ditto waves extending beyond the mean-N IGW-bounds. At the low-frequency, sub-inertial side of the IGW-band the rare $N_{min}$ may combine with sub-mesoscale motions. More importantly, at the high-frequency, super-buoyancy side $N_s$ may combine with turbulent overturns. If all relevant scales are resolved, a safe separation frequency would thus be at overall maximum $N_{max} = max(max(N_s))$, where the maximum between brackets is determined for each profile. In practice, such a transition frequency between internal waves and turbulence is not easily determined, because it requires the small scales to resolve relevant $L_O$. Only in weakly stratified waters, $L_O$ are O(10) m, and resolution of 1-2 m scales should be sufficient.

Thus, under weakly stratified conditions, instrumental noise and short-term bias have to be corrected in t- and z-direction, respectively. A practical solution that also eliminates internal waves and sub-mesoscale motions, is application of a band-pass filter 'bpf' in t, an lpf and sorting in z so that per time step (A2) reads,

$|\theta'_{bpf}|/(d\theta_{sorted}/dz)$,

to which sufficient averaging is applied. Noting that stratification varies over different scales by two orders of magnitude, so that $N_s \approx 0$-6f, filter design discriminates between conditions of SW and NH. Sharp, phase-preserving double-elliptic filters (Parks and Burrus, 1987) are designed following inspection of temperature variance spectra (Fig. A2).

For SW, reasonable filter cut-offs are tuned for a 1.7-day period around day 308 by equating (A3) to $\varepsilon_T$. SW's lpf cut-off is fixed at 3000 cpd. The reference hpf cut-off frequency $\omega_{hpf,ref}$ appeared at a small flat (0-)slope near the low end of turbulence buoyancy-($\omega^{-7/5}$) and inertial-subrange ($\omega^{-5/3}$) slopes, where a short steep slope to internal-wave frequencies occurred. For other SW-periods, reference is not made using average large-scale N, but a better fit is found for the time average of maximum small-scale buoyancy frequencies per profile $N_m = <max(N_s)>$ so that,

$\omega^{SW}_{hpf} = (N_m/N_{m,ref})^2\omega_{hpf,ref}$. (A4)

For NH, the lpf cut-off is fixed at 500 cpd. For its hpf cut-off, a flat (0-) slope appeared at a frequency just higher than $\omega_{max}$ so that, independent of measured $<N_{s,max}>$ the cut-off is blocked at,

$\omega^{NH}_{hpf} = 3.7$ cpd. (A5)

This value is tuned for a period with dominant GH, for which the geophysics-determined (e.g., Pasquale et al., 1996) buoyancy flux $fl/\Gamma_C = \varepsilon_{GH} = 1.2\times10^{-10}$ m$^2$ s$^{-3}$. In the present (Fig. 6) and previous (van Haren, 2025) data the mixing coefficient was found to amount $\Gamma_C = 0.5$, which is typical for convection turbulence (Dalziel et al., 2008; Ng et al., 2016).

The cut-off frequency in (A5) is to within $\pm 0.2$ cpd equivalent to $1.8(2\Omega) \approx 1.8<\omega_{max}> \approx 2<N_{s,max}>$ $\approx 0.5<U>/<L_T> = 0.5\omega_O$, for the deep Western Mediterranean site. $\Omega$ denotes the Earth rotation, and U the waterflow speed. The Ozmidov (1965) frequency $\omega_O$ is a natural separator between internal waves and stratified turbulence. It is unknown why the filter cut-off is close to half the Ozmidov frequency. Also puzzling is the lack of correspondence between (A4) and $\omega_O$, with the factor varying between 0.3 and 0.9 for different periods of SW. In part this may have to do with the waterflow being measured at h $= 126$ m, below which it may be more uniform under NH- than under SW-conditions.

As shown in Fig. 4 for (A4), and Fig. 5 for (A5), the comparison works well to within about 20%. For Fig. 4, $<[\varepsilon_T]> = 6.3\times10^{-10}$ m$^2$ s$^{-3}$ and $<[\varepsilon_E]> = 6.7\times10^{-10}$ m$^2$ s$^{-3}$. For Fig. 5, $<[\varepsilon_T]> = 2.0\times10^{-10}$ m$^2$ s$^{-3}$ and $<[\varepsilon_E]> = 1.3\times10^{-10}$ m$^2$ s$^{-3}$. Further tests were performed for about ten 1-4 day periods of NH and SW, and all were within above error, provided that the data post-processing was carefully done and longer periods were avoided. Especially (A5) is very sensitive to small changes in filter steepness around the cut-off frequency, presumably due to its proximity to the IGW upper bound.

**Table 1.** General first and second moment statistics calculated for 45-line, 124-m height, and half-day period turbulence dissipation rate values of Fig. 6. Two environmental conditions are characterized: SW = stratified-water, NH = near-homogenous, and may include GH. The two conditions are separated by a criterion equivalent of N = 0.65f.

| Day | $\varepsilon_T \ [m^2 \ s^{-3}]$ | Cond. |
|-----|-----|-----|
| 308 | $4.0\pm0.8\times10^{-10}$ | SW |
| 441 | $1.5\pm0.1\times10^{-9}$ | SW |
| 459 | $1.0\pm0.2\times10^{-10}$ | NH |
| 495 | $1.0\pm0.2\times10^{-10}$ | NH |
| 485 | $4.8\pm0.4\times10^{-10}$ | SW |
| 652 | $1.4\pm0.1\times10^{-10}$ | NH |

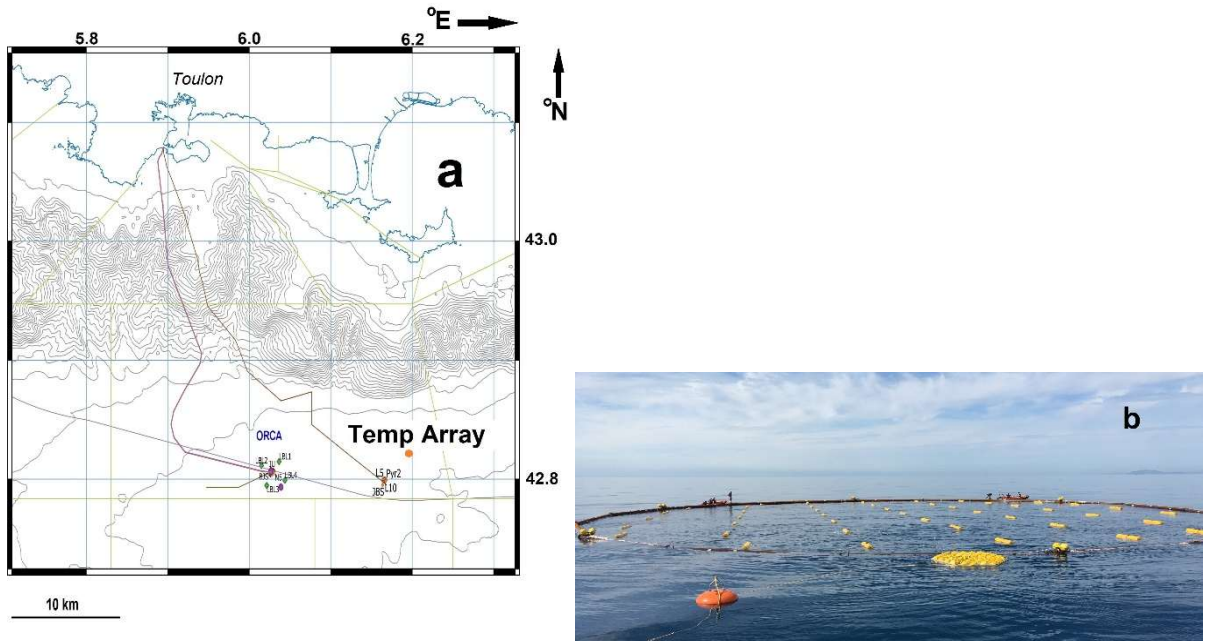

**Figure 1.** Large-ring mooring site and deployment. (a) Location named "Temp Array" (orange dot) on map off southern France. The mooring is well east of main neutrino telescope 'NT' site "ORCA" of KM3NeT (Adrián-Martinez et al., 2016) and just northeast of the former ANTARES NT-site. Isobaths are drawn every 100 m. (b) At sea, during deployment finalizing the opening of air valves before sinking. The near part of the large steel-tube ring is already underwater. Almost all buoys of the 45 small-ring compacted vertical lines are visible (for layout see Appendix A1).

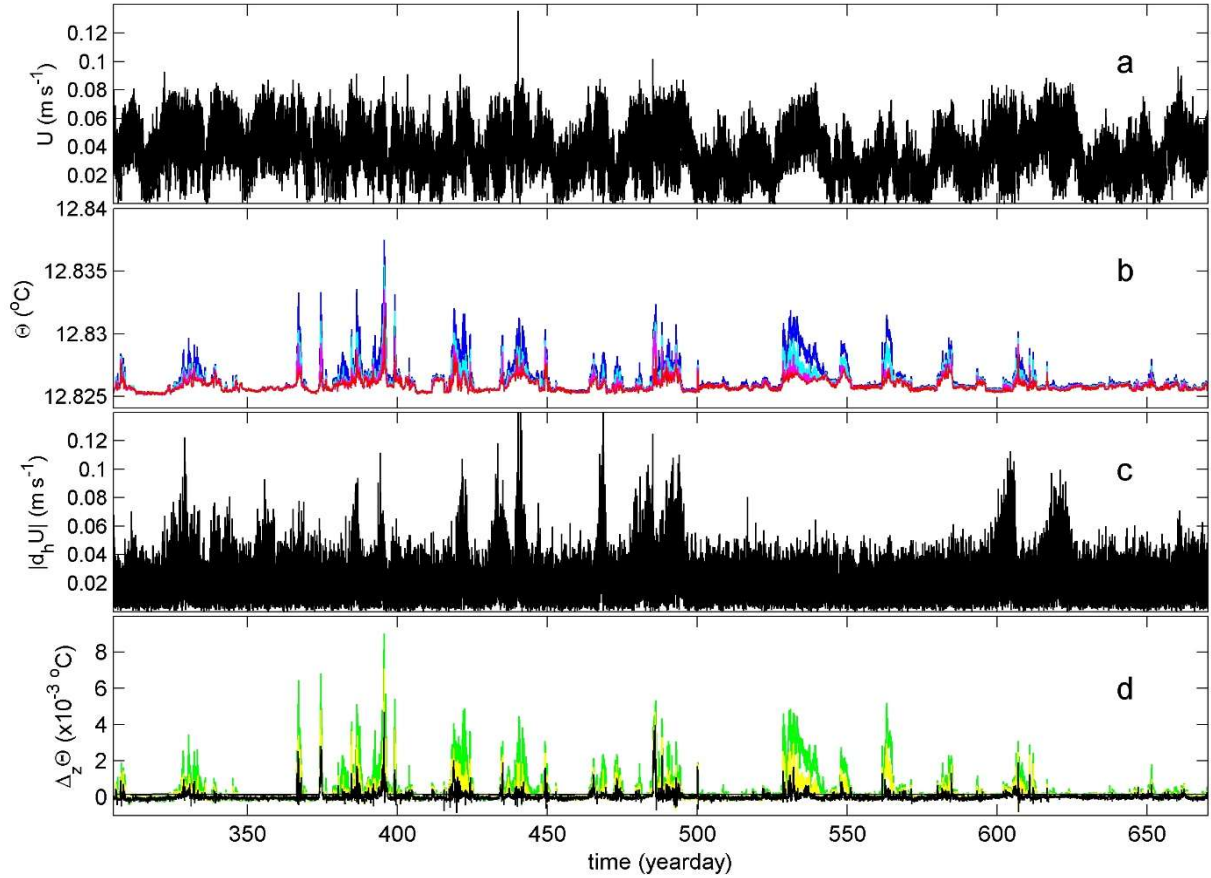

**Figure 2.** Time series of T-sensor and current meter 'CM' data, for the first year after deployment. Time in days of year 2020, +366 in 2021. (a) Unfiltered waterflow speed at h = 126 m above seafloor. (b) Conservative Temperature from vertical line 25, at h = 1.5 (red), 29.5 (magenta), 59.5 (cyan) and 99.5 m (blue), corrected for drift and referenced to CTD-data of Fig. 3b. (c) Amplitude of horizontal flow differences. (d) Vertical temperature differences between the lowest T-sensor and those above from b.

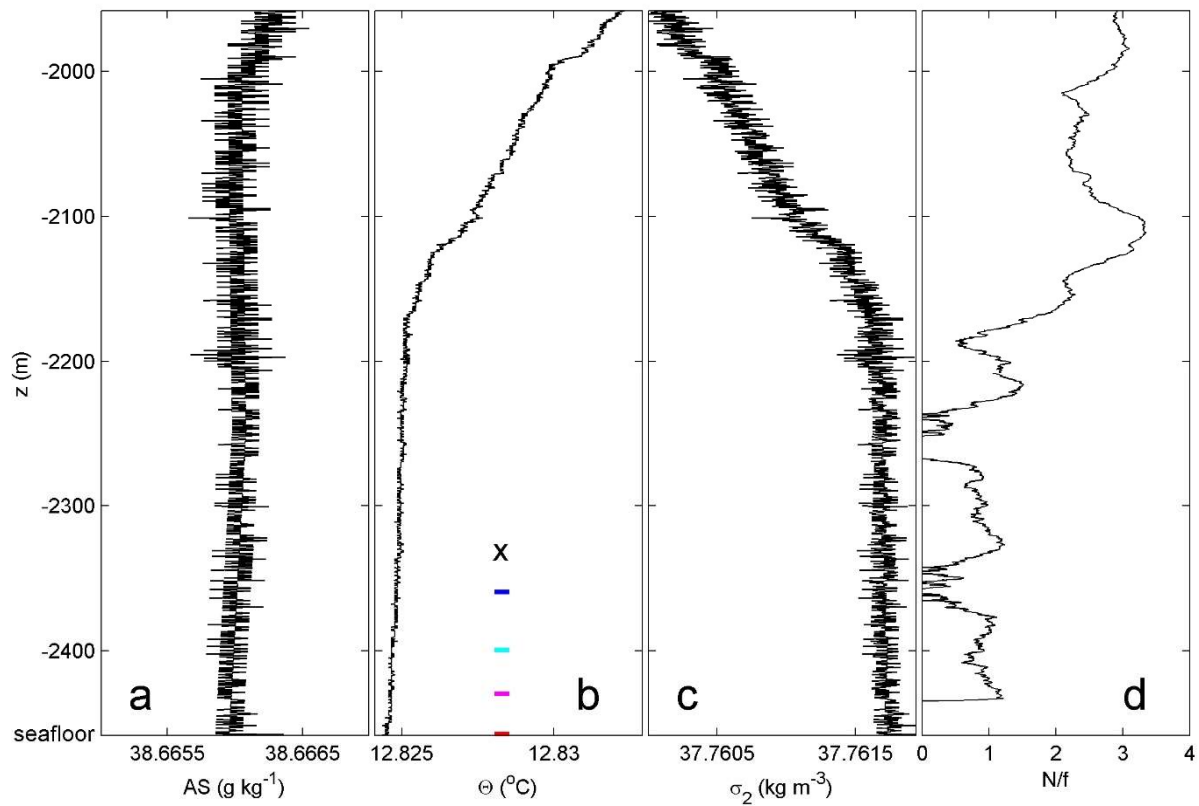

**Figure 3.** Lower 500 m of shipborne CTD profile obtained to within 0.5 m from the seafloor (at z = -2458 m). (a) Absolute Salinity, with x-axis range similar to that of b. in terms of contribution to density variations. (b) Conservative Temperature. The colored ticks indicate the vertical positions of the four T-sensors of which data are displayed in Fig. 2a. The 'x' indicates the position of CM. (c) Density anomaly referenced to $2×10^7$ N m$^{-2}$. (d) Ratio of 25-m scale buoyancy frequency 'N' over local inertial frequency 'f'.

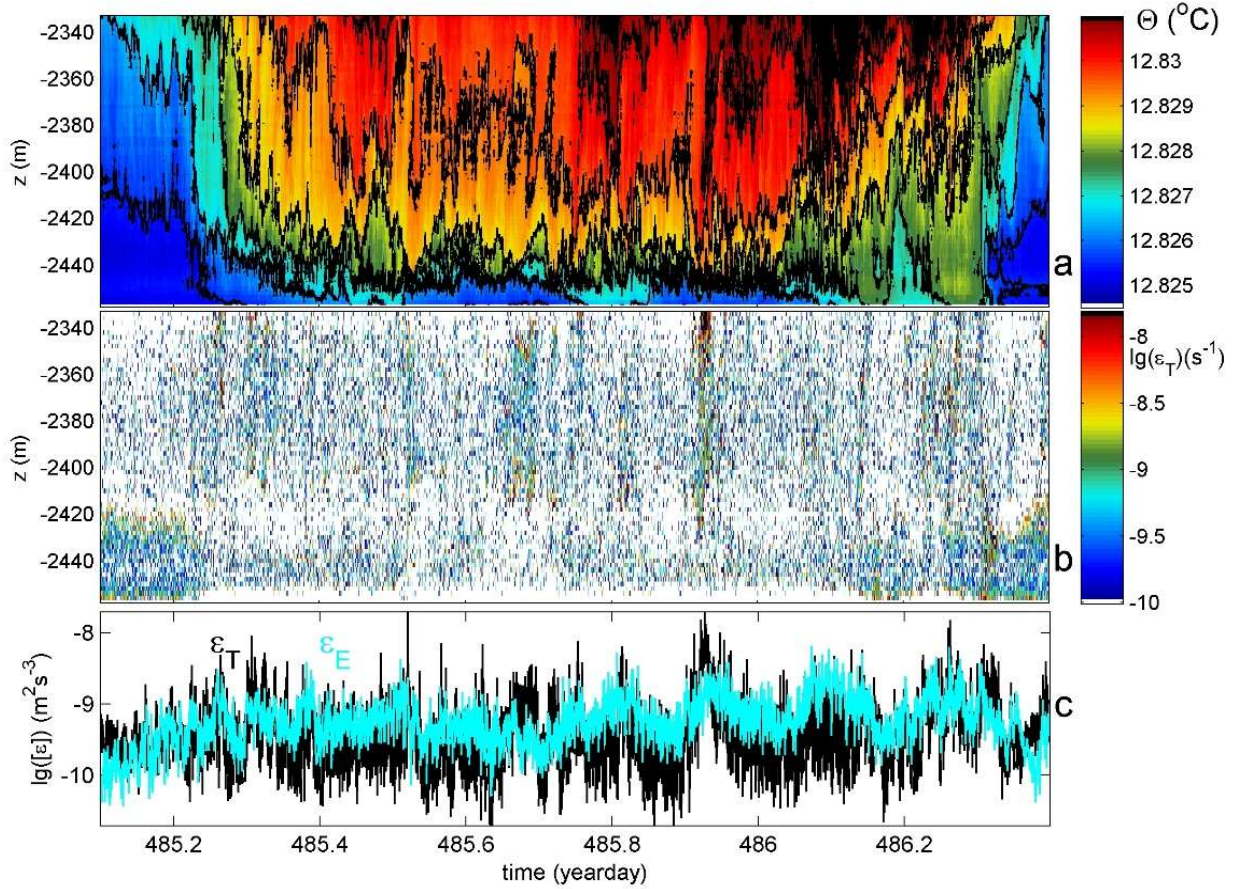

**Figure 4.** A 1.3-day period of relatively strong stratification with maximum small-scale buoyancy frequency $N_{max}$ = 6f, for data from vertical line 15. (a) Time-depth plot of Conservative Temperature with black contours every 0.001°C. The horizontal axis is at the seafloor. (b) Logarithm of non-averaged turbulence dissipation rate from data in a. using Thorpe (1977) method. (c) Time series of logarithm of data from b. averaged over 124-m vertical extent of T-sensors (black), compared with calculations using Ellison (1957) method (cyan) with high-pass filter 'hpf' cut-off from Fig. A2a.

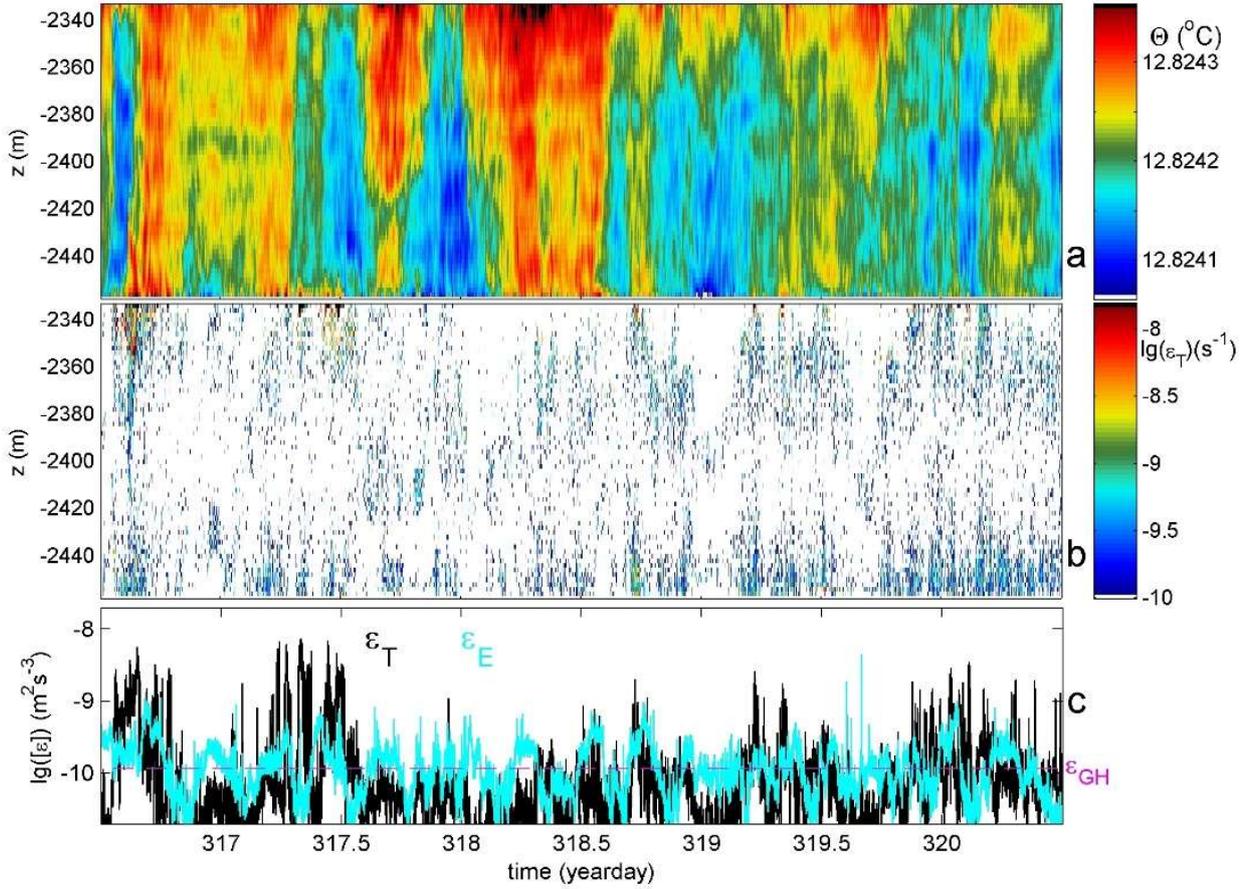

**Figure 5**. As Fig. 4, but for four days of near-homogeneous conditions with mean N ≈ 0.5f, very weak stratification alternated with convectively unstable periods. For c., the hpf cut-off for determining $\varepsilon_E$ is shown in Fig. A2b and the magenta-dashed line indicates the average dissipation rate attributed to geothermal heating 'GH' (see text).

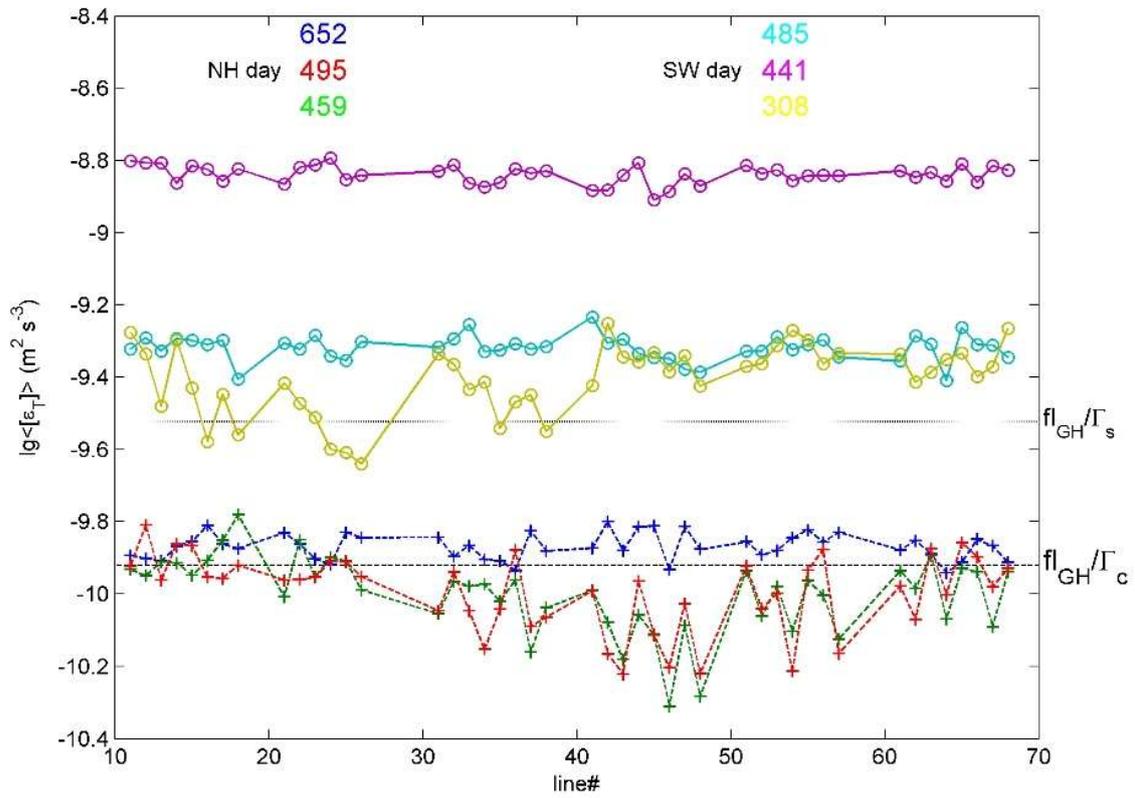

**Figure 6.** Limited statistics of Thorpe (1977) method (logarithm of) turbulence dissipation rates averaged over 124 m vertically given for six 12-h periods indicated by day-number, as a function of all 45 lines that are indicated their number 'line#'. Two thresholds are given as a function of general average buoyancy flux 'fl' from GH, divided by mixing coefficient for convection-turbulence $\Gamma_c = 0.5$ (Dalziel et al., 2008) and for shear-turbulence $\Gamma_s = 0.2$ (Osborn, 1980; Oakey, 1982). Solid lines (o) indicate Stratified-Water 'SW' conditions, dashed lines (+) indicate near-homogeneous 'NH' conditions.

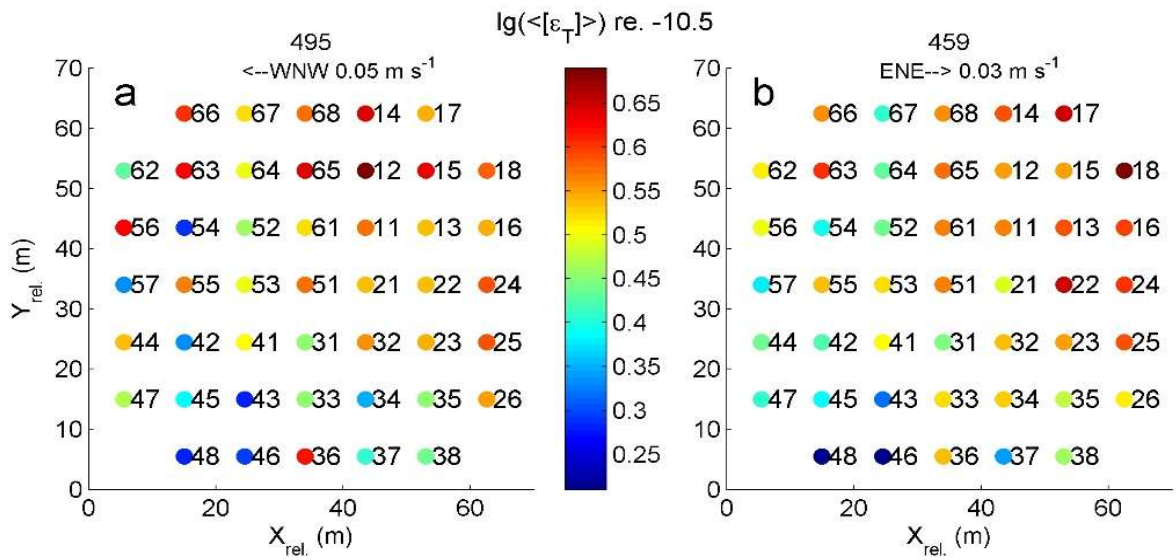

**Figure 7.** Plan view of 4 lines indicating logarithm relative to a value of -10.5 of time- and vertical-mean turbulence dissipation rates for NH conditions on days 459 and 495 of Fig. 6. On top, half-day mean waterflows are indicated.

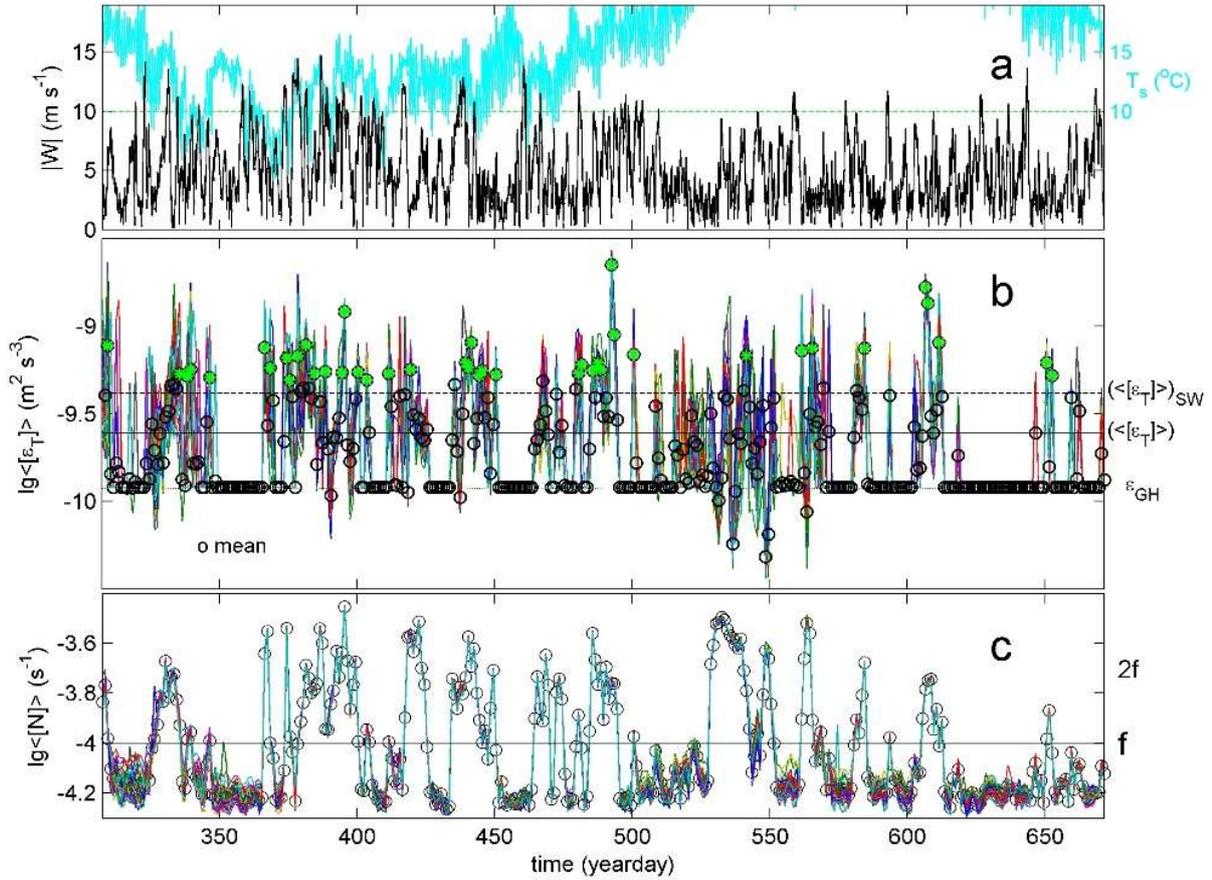

**Figure 8.** Yearlong time series of 45-line, daily, and vertically averaged turbulence and stratification values compared with meteorological data. (a) Wind speed (black) and surface temperature (cyan) measured at the station of Porquerolles Island, 20 km north of the mooring-array. The horizontal line is an arbitrary reference line below which near-surface convection may occur under sufficient pre-conditioning. (b) Logarithm of daily and 124-m vertically averaged Thorpe (1977) method turbulence dissipation rate for all 45 lines (colour), including their mean values (black, circles) of which those exceeding twice the overall mean value (green asterisks). A threshold of 0.0002°C is applied for T(125)-T(1), below which values are forced to mean $\varepsilon_{GH} = 1.2 \times 10^{-10}$ m$^2$ s$^{-3}$, see text. The solid horizontal line indicates the overall mean value (4), the dashed line the mean (5) for periods under SW conditions. (c) Logarithm of corresponding mean buoyancy frequencies from reordered temperature profiles. The horizontal line indicates the local planetary inertial frequency.

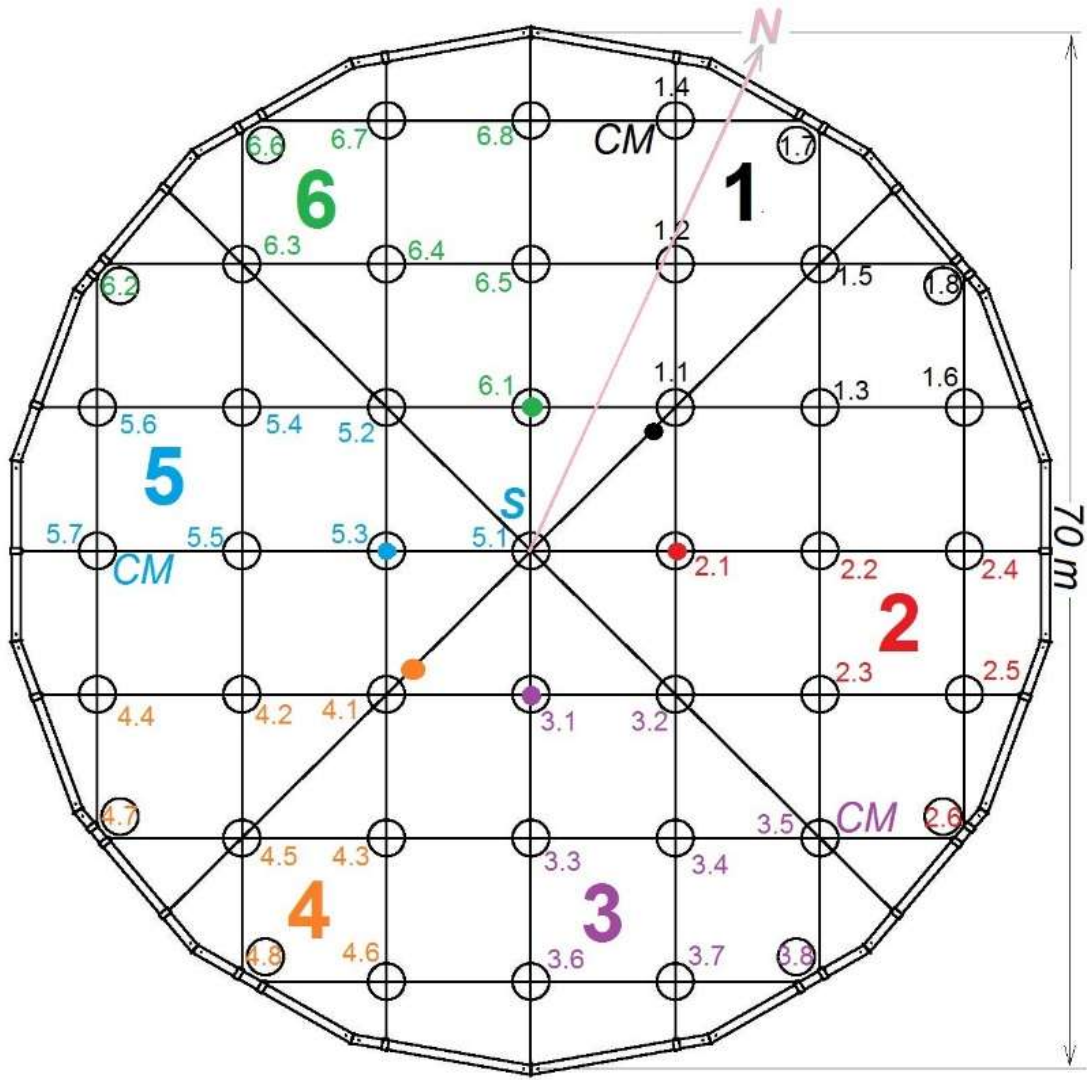

**Figure A1.** Orientation and layout of the large-ring mooring, with steel-cable grid and small rings numbered in six synchronisation groups with colour dots indicating group nodes and synchronier 'S' at line 51. Here and elsewhere in the text, lines are indicated without period for short. Lines 14, 35 and 57 held a waterflow current meter 'CM' at the buoy.

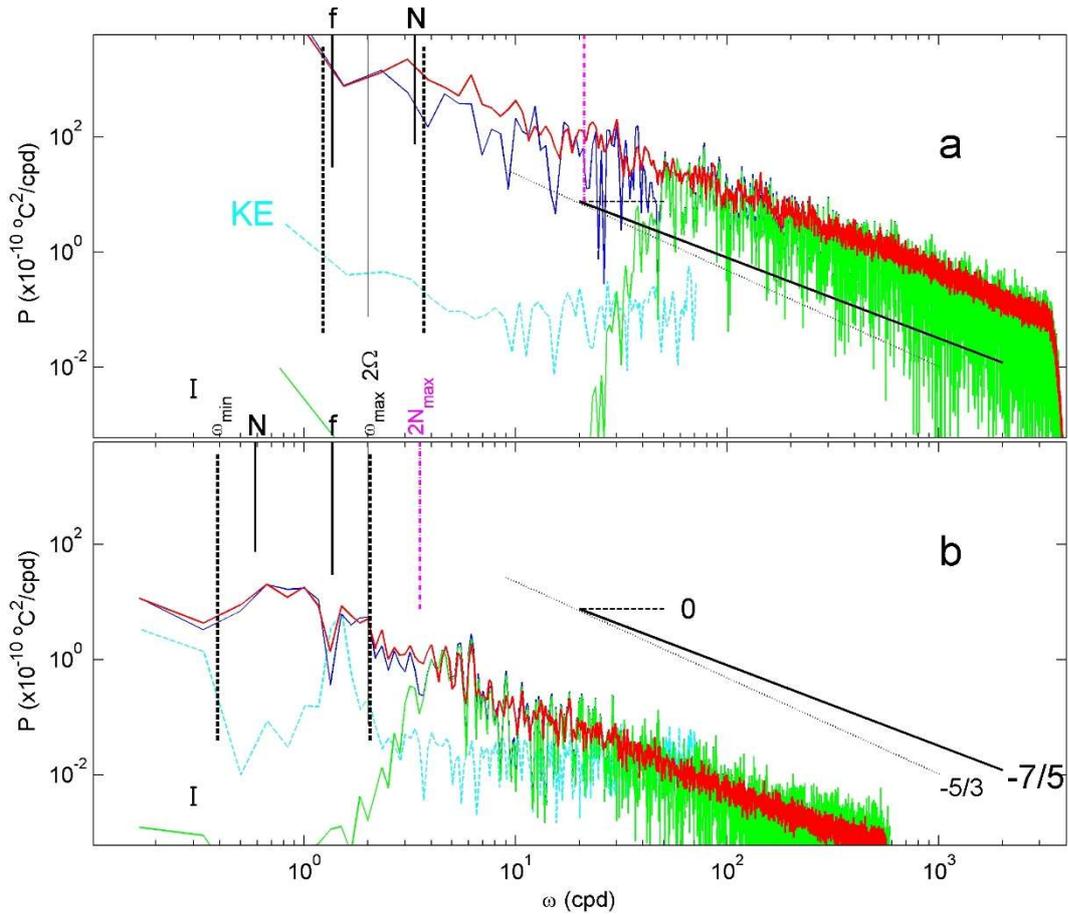

**Figure A2.** Spectra demonstrating filter cut-off frequencies for Ellison (1957) method under SW and NH conditions. Data from line 15. Weakly smoothed (6 degreed of freedom, dof) low-pass filtered 'lpf' spectra from a single T-sensor at mid-height (blue) with hpf version (green) is compared with moderately smoothed (100-dof) lpf spectrum over all 63 T-sensors (red). For comparison, the weakly smoothed (10-dof) kinetic energy 'KE' spectra are given (cyan; arbitrary vertical scale), averaged over the three CM. For reference, several frequencies and turbulence-range spectral slopes are given, see text. (a) SW period of Fig. 4, with filter cut-off following a scaling of time mean $<N_{max}>^2$ and lpf cut-off at 3000 cpd (cycles per day). (b) NH period of Fig. 5, with hpf cut-off fixed near $2<N_{max}>$ and lpf cut-off at 500 cpd.